# A Framework for Modelling Probabilistic Uncertainty in Rainfall Scenario Analysis

*Seyed Hamed Alemohammad [1], Reza Ardakanian [2] and Akbar Karimi [3]*

[1] *Graduate Student of Water Resources Engineering, Department of Civil Engineering,
Sharif University of Technology, Tehran, 11155-9313, Iran, e-mail: alemohammad@civil.sharif.edu*
[2] *Assistant Professor, Department of Civil Engineering,
Sharif University of Technology, Tehran, 11155-9313, Iran, e-mail: ardakanian@civil.sharif.edu*
[3] *Ph.D. Candidate of Water Resources Management, Department of Civil Engineering,
Sharif University of Technology, Tehran, 11155-9313, Iran, e-mail: ak_karimi@civil.sharif.edu*

*Abstract*

*Predicting future probable values of model parameters, is an essential pre-requisite for assessing model decision reliability in an uncertain environment. Scenario Analysis is a methodology for modelling uncertainty in water resources management modelling. Uncertainty if not considered appropriately in decision making will decrease reliability of decisions, especially in long-term planning. One of the challenges in Scenario Analysis is how scenarios are made. One of the most approved methods is statistical modelling based on Auto-Regressive models. Stream flow future scenarios in developed basins that human has made changes to the natural flow process could not be generated directly by ARMA modelling. In this case, making scenarios for monthly rainfall and using it in a water resources system model makes more sense. Rainfall is an ephemeral process which has zero values in some months which introduces some limitations in making use of monthly ARMA model. Therefore, a two stage modelling approach is adopted here which in the first stage yearly modelling is done. Within this yearly model three ranges are identified: Dry, Normal and Wet. In the normal range yearly ARMA modelling is used. Dry and Wet range are considered as random processes and are modelled by frequency analysis. Monthly distribution of rainfall, which is extracted from available data from a moving average are considered to be deterministic and fixed in time. Each rainfall scenario is composed of a yearly ARMA process super-imposed by dry and wet events according to the frequency analysis. This modelling framework is applied to available data from three rain-gauge stations in Iran. Results show this modelling approach has better consistency with observed data in comparison with making use of ARMA modelling alone.*

## 1. INTRODUCTION

Predicting future probable values of model parameters, is an essential pre-requisite for assessing model decision reliability. Scenario Analysis is a methodology for modelling uncertainty in water resources management modelling. Uncertainty if not considered appropriately in decision making will decrease reliability of decisions, especially in long-term planning. One of the challenges in Scenario Analysis is how scenarios are made. One of the most approved methods is statistical modelling based on Auto-Regressive models. Auto-Regressive modelling described flow processes in cases where extreme values are not present (zeroes, very high/very low flows). Burlando *et* al (1993) in a paper have used the ARMA model to forecast short time hourly precipitation. But they haven't considered the influence of the storm movement. Two models have been used: the first uses the data which are collected continually during the month or season and the second adaptively uses the data of a specific flood, which results better than the first one. Kendall & Dracup (1991) have compared the two models AR (1) and Index-sequential, for modelling runoff. A 32 years runoff the model has shown that AR (1) can't forecast the reservoir storage with good approximation. But for a yearly model the results are to some extents the same. Chebaane & Salas (1995) have described a method to solve the problem of modelling zero in time series. The model assumes that the monthly discharge is the product of an alternative discrete binary process and an alternative continuous binary process, which both are a first-order autoregressive process. The discrete process shows the occurrence or not occurrence of discharge and the continuous one gives the discharge at the time that it occurred. Lunge & Sefe (1991) have modelled monthly runoff using a stochastic model. Total monthly runoff sequences behaved as an integrated moving average process (IMA (0, 1, 2)). The model output for a 12 month forecast has been satisfactory.

Hsu & Adamowski (1981) have presented a model for forecasting river discharge using meteorological data as input. A First-order autoregressive moving average process has been used as a transfer function for runoff time series and meteorological data. The weakness of the model would be lack of meteorological data. In this research a framework is presented which divides the time-series to normal, wet and dry sets. Normal set is modelled by ARMA process but wet and dry sets by frequency analysis. In modelling and generating scenarios according to the length of samples (scenarios) and frequency of occurrence of dry and wet flows, they are super-imposed in generated scenarios by ARMA.

## 2. MODELLING FRAMEWORK

Considering the components of a time series, a structure as below is assumed for the model:

1. Producing yearly series by summing monthly series;
2. Determining the monthly distribution;
3. Modelling the yearly series using ARMA(p,q) and generating data;
4. Finding the monthly distribution of the yearly generated series.

As the input data are monthly precipitations, they should be converted to yearly precipitations by adding up each 12 consequent data.

$$Y_Q(y) = \sum_{m=1}^{12} M_{Q_{m,y}}$$

Where:

$Y_Q(y)$ = yearly precipitation

$M_{Q_{m,y}}$ = monthly precipitation

In the next step the monthly distribution should be found. Then an ARMA modelling will be executed to see whether a (p, q) can be found to model the data or not? If a (p, q) found then data will be sent to the fourth part of the model to find the monthly distribution of the generated data.

$$Y_Q(y) = ARMA(p,q)$$

But if the appropriate (p, q) could not be found a categorization of the precipitations to wet, normal and dry should be done. To find a normal limit and then the wet and dry limit, an iterative procedure should be done. By defining the normal limit a range for normal data will be determined and the data that are in this range will be used for modelling the yearly series. This classification of precipitations is to eliminate the random precipitations from the normal ones for better modelling of the flow. In the normal set, the data that are above the normal limit will be called normal up and others normal down. Two other sets will be determined: Wet and Dry, that each of them will be categorized to two subsets: very wet and wet, very dry and dry. The very dry subset is the zero data. After the categorization, each year should be checked to find its subset and then the frequency of each type of subsets should be determined. This calculation is done to be able to substitute the dry and wet years among the normal years after modelling the yearly series. Finishing this part, the yearly series is ready and using the multipliers of the second part the monthly precipitations can be found. The assumptions made in this modelling were:

1. Future is like past and frequency of wet and dry years will be like the past.
2. The yearly distribution of data is an ARMA process.
3. The monthly distribution of data is deterministic and not related to the yearly amount of rainfall.

4. There isn't a trend in the input precipitations. It means that there won't be a main change in the nature to change the process of rain and if something exists it's filtered before using the model.

5. The distribution of the extreme values (dry and wet sets) is different from the normal values and they are modelled using frequency analysis.

6. Number of the input precipitations is a multiple of 12. It means that we have all the monthly data of the period of study and no data is missing.

The flow chart of the model is shown in figure 1.

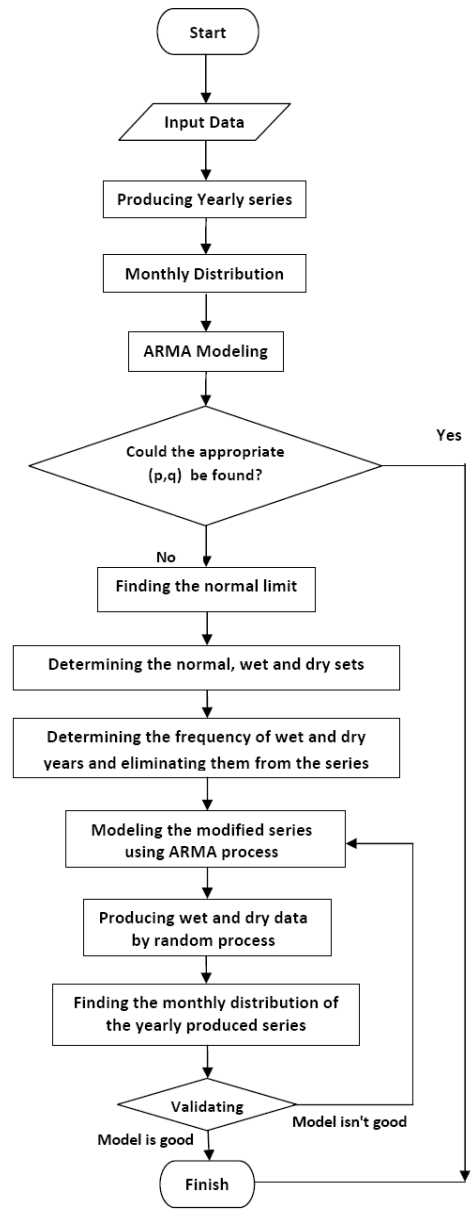

**Figure 1. Model Flow Chart**

## 3. MODELLNG PROCEDURE

In this part according to the structure described in the last part the procedure of modelling will be discussed:

- **Producing yearly series using monthly series;**

    This part is a simple calculation and just each 12 data of monthly series will be summed to determine the yearly series.

- **Determining the monthly distribution;**

    To do this a sum of the ratios of the precipitation in each month to the precipitation of the year, for different years will be divided by the number of years to find the factor of the specific month.

$$Snf_m = \frac{1}{N} \sum_{y=1}^{N} \left[ \frac{M_{Q_{m,y}}}{Y_Q(y)} \right]$$

Where:

$Snf_m$ = the seasonality factor which is the month's factor

$N$ = the number of years in the input data

- **Modelling the yearly series using ARMA(p,q);**

    This section is the main part of modelling and is made of seven parts:

    1) Modelling using ARMA(p,q)

        Yearly series should be modelled to see if any ARMA (p, q) could predict its behaviour. If the appropriate (p, q) could be found, the modelling will go to the last section and the monthly precipitations are found. The criteria for choosing the best (p, q) are the AIC (Akaike Information Criterion) and FPE (Final Prediction Error) indexes. If the appropriate (p, q) couldn't be found the procedure will be continued to the next step.

    2) Finding the normal limit

        The normal limit is the real average of the yearly data but in the most data there are some very dry or very wet rainfalls that deviate this limit. To reduce the effect of these data on the normal limit an iterative process will be executed. First the average of the data will be found and then the data that are out of the 80% central part of the domain of data are eliminated and again the mean of the new data will be found. This procedure will be continued till one of the below conditions is satisfied:

        I. The new mean doesn't change to more than 10% of the last mean;

        II. The number of the data becomes less than 50% of the total data.

        After this procedure the normal limit is determined.

    3) Defining the normal, wet and dry ranges

        The normal set is the range of the data used to find the normal limit of the precipitations. Therefore the wet set is from the up boundary of the normal limit to the maximum precipitation and the dry set is from the lower boundary of the normal set to the minimum precipitation.

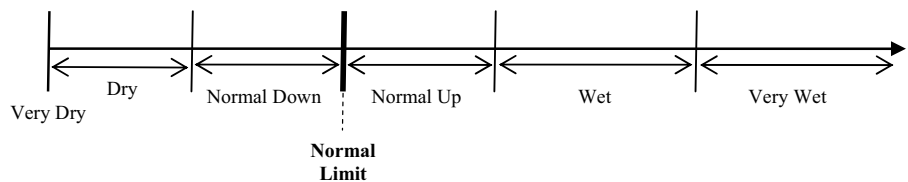

Figure 2. The Six Sets of Rainfall

It's worth mentioning that the very dry subset is just the zero data and is not a range. The reason of this type of categorization is because of the importance of the years with zero precipitations (droughts).

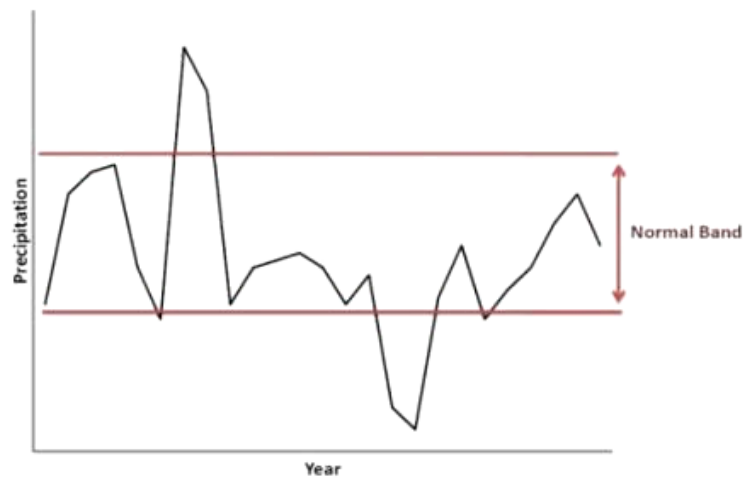

**Figure 3. The Normal Band**

4) Finding the dry and wet years and their frequency

   Using the limits defined in the last step now it should be determined that each year is located in which set. By this classification the frequency of the wet and dry years (also their subsets) will be determined that will be used in producing the final yearly series.

5) Eliminating the dry and wet years from the yearly series

   Before modelling the yearly series the wet and dry data should be eliminated from the data series. Indeed a filtering is done to eliminate the cyclic component of the series if exist. The final series that is derived by eliminating the wet and dry data is called the modified series.

6) Modelling the modified series using ARMA process

   This step is done according to the second assumption of the model. The best (p, q) for modelling the modified series will be found using the same criteria that was used before: AIC and FPE.

   After finding the appropriate (p, q) a series of data with the same number of years as the input series will be produced.

7) Producing wet and dry data by random process

   In the third step the frequency of the dry and wet years was determined and now for each occurrence of these cases two random numbers will be produced: the first in the range of the number of the years to determine the year that the specific case occurs, and the second in the range of the related set to determine the amount of precipitation. These will be substituted in the produced series of the last step to define the final yearly series.

- **Finding the monthly distribution of the yearly produced series.**

    Now by having the yearly series and the monthly distribution of precipitations the monthly series can be determined and the modelling is finished.

- **Producing Scenarios;**

    The goal of this modelling was to define the criteria for modelling uncertainty in scenario analysis. In this step the scenarios can be made. After the determination of (p, q) the procedure can be repeated for any times (the same as the number of the scenarios needed) to produce different sets of yearly data and finally different sets of monthly flows.

## 4. CASE STUDIES

To verify the developed modelling framework three real world cases are modelled and their results are compared with observed values. These three real cases are stations selected from different parts in Iran: Amirkabir Dam station, Babol Station and Diva Station. The results of these modelling are shown in the following charts.

The Amirkabir Dam station is located in North Mountains of Iran with an average of precipitation about 405 mm per year. The Babol Station is in Northern part of Iran near the Caspian Sea with an average of rainfall of 662.8 mm per year. The Diva Station is also near the Caspian Sea with an average of rainfall equal to 882.75 mm per year. For the Amirkabir Dam Station there were 29 years of data, for Babol Station there were 33 years of data and for Diva Station there were 21 years of data. Regarding the ARMA modelling of the normal series in different stations, it should be mentioned that for Amirkabir Dam the best (p, q) were found to be (0, 0) but for Babol it was (2, 1) and for Diva it was (2, 3).

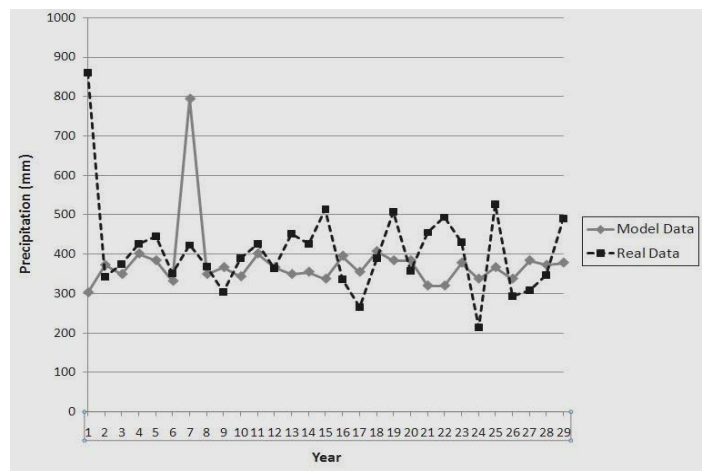

**Figure 4. Comparison of Real and Model Yearly series for Amirkabir Dam Station**

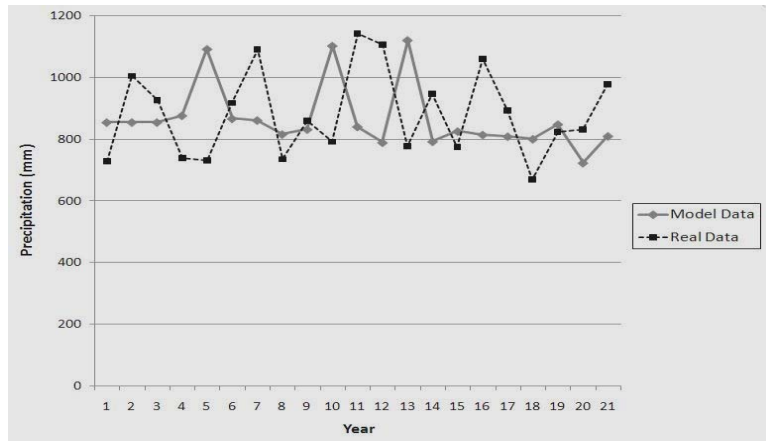

**Figure 5. Comparison of Real and Model Yearly series for Diva Station**

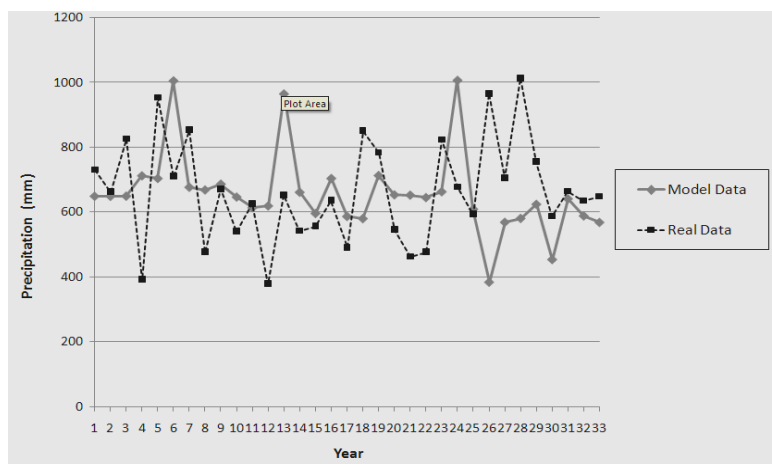

**Figure 6. Comparison of Real and Model Yearly series for Babol Station**

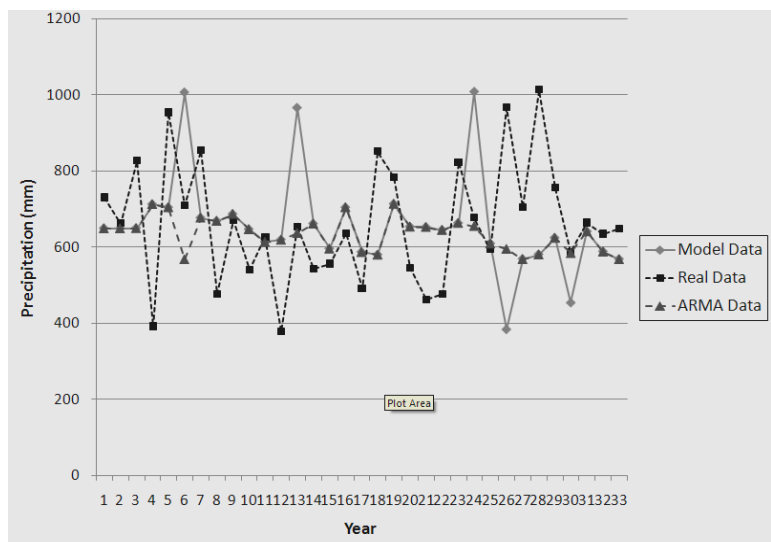

**Figure 7. Comparison of Real, Model and ARMA Yearly series for Babol Station**

Also it can be seen that the model results are better than ARMA modelling results (Fig. 7). The results show that the mean and variance of the real data and model data are similar, as criteria for model assessment, and this affirms the use of the model.

## 5. CONCLUSIONS

Results showed that using this modelling framework for producing scenarios will make better results and accordance with reality in comparison with just using ARMA (p, q) process. It was also shown that in a prediction range this model can predict wet and dry years more accurate than the ARMA process alone. It should be mentioned again that the procedure assumed for the wet and dry years was a pure random process and it's a reasonable assumption. Because if they weren't random, then were correlated to last years of occurrence and would be modelled in the ARMA process. Finally it can be said that, this modelling framework works as a more general approach to time-series modelling, with extreme values. This modelling framework helps to better scenario making which represents rainfall process.